\documentclass[10pt,prl,twocolumn,superscriptaddress,showpacs,floatfix]{revtex4}

\usepackage{graphicx}
\usepackage{amssymb}
\usepackage{times}
\usepackage{amsmath}
\usepackage{amsthm}

\newcommand{\ket}[1]{|#1\rangle}

\begin{document}

\title{Quantum Cryptography Approaching the Classical Limit}

\author{Christian Weedbrook}\email{christian.weedbrook@gmail.com} \affiliation{Department of Physics, University
of Queensland, St Lucia, Queensland 4072, Australia} \affiliation{Research Laboratory of Electronics, Massachusetts Institute of Technology, Cambridge MA 02139, USA}

\author{Stefano Pirandola} \affiliation{Department of Computer Science, University of York, York YO10 5DD, United Kingdom}

\author{Seth Lloyd} \affiliation{Research Laboratory of Electronics, Massachusetts Institute of Technology, Cambridge MA 02139, USA}\affiliation{Department of Mechanical Engineering, Massachusetts Institute of Technology, Cambridge MA 02139, USA}

\author{Timothy C. Ralph} \affiliation{Department of Physics, University
of Queensland, St Lucia, Queensland 4072, Australia}

\date{\today}

\begin{abstract}

We consider the security of continuous-variable quantum cryptography as we approach the classical-limit, i.e., when the unknown preparation noise at the sender's station becomes significantly noisy or thermal (even by as much as $10^4$ times greater than the variance of the vacuum mode). We show that, provided the channel transmission losses do not exceed $50\%$, the security of quantum cryptography is not dependent on the channel transmission, and is therefore, incredibly robust against significant amounts of excess preparation noise. We extend these results to consider for the first time quantum cryptography at wavelengths considerably longer than optical and find that regions of security still exist all the way down to the microwave.

\end{abstract}


\maketitle

\textbf{{\it Introduction}} - Quantum key distribution (QKD) using continuous variables~(CV)~\cite{Cer07,Sca09} allows two people, Alice and Bob, to generate a secure key which can be used to encrypt messages. CV-QKD protocols using Gaussian modulation~\cite{F.Grosshans2002,F.Grosshans2003,Sil02,C.Weedbrook2004,Lan05,S.Pirandola2008}, initially begin with Alice preparing a number of randomly displaced pure coherent states and sending them over an insecure quantum channel to Bob. Generally, it is assumed that Alice's states must be pure quantum states to a good approximation otherwise her ability to perform QKD will rapidly become compromised. This seemed to be borne out by recent calculations~\cite{Fil08} that showed that the distance over which CV-QKD was secure, when Alice used mixed coherent states in the protocol, fell rapidly as the states became significantly impure.


In this Letter, we show that, provided the channel transmission losses do not exceed $50~\%$, the security of quantum cryptography is not dependent on the channel transmission, and is therefore incredibly robust against significant levels of impurity of Alice's states, without the additional previous requirement of purifiers~\cite{Fil08}.
%
%
This is a remarkable result as we might naturally expect that as Alice's states become more and more thermalized secure transmission over any finite distance would become impossible. This further motivates an investigation of the security of CV-QKD as we move from optical frequencies into the infrared and down into the microwave region. As the wavelength gets longer there is no direct way of detecting single photons~\cite{Tem06} thus ruling out discrete variable approaches. While CV measurements still apply, state preparation and the quantum channel become thermalized by the significant levels of background radiation that exist for longer wavelengths at room temperature. Here we show that CV-QKD remains, in principle, possible over short distances, well into the infrared and into the microwave regime. This surprising result highlights the possibility of short-range quantum cryptography applications at sub-optical frequencies.

\textbf{{\it Quantum Cryptography using Gaussian States}} - Typical Gaussian modulated CV-QKD protocols, begin with Alice randomly modulating a vacuum state to create a coherent state $\ket{\alpha}$~\cite{Ger05}. This random modulation or displacement $\alpha = Q_A+iP_A$ contains two independent variables $X_S \in \{Q_A, P_A\}$ chosen from a two-dimensional Gaussian distribution with variance $V_S$ and zero mean. It is these continuous variables that will ultimately be used to construct a secret key between Alice and Bob. Alice then sends a whole ensemble of these randomly displaced pure coherent states to Bob over a quantum channel which is monitored by the eavesdropper, Eve. At the output of the channel, Bob measures the incoming states using either homodyne~\cite{F.Grosshans2003} or heterodyne detection~\cite{C.Weedbrook2004}.

The initial modes prepared by Alice can be described in the Heisenberg picture as $\hat{X}_A = X_S + \hat{X}_0$ where $X_S$ describes the classical signal and $\hat{X}_0$ the thermal mode. Here the quadratures $\hat{Q}$ and $\hat{P}$ are defined as: $\hat{X}_A \in \{\hat{Q}_A, \hat{P}_A\}$ and $\hat{X}_0 \in \{\hat{Q}_0, \hat{P}_0\}$. The overall variance $V := V(\hat{X}_A)$ of Alice's initially prepared mode is given by: $V = V_S + V_0$. We can further decompose the variance of the thermal mode $V_0:=V(\hat{X}_0)$ into the variance of the pure vacuum mode (which is normalized to 1) and the variance of the unknown preparation noise at Alice's station $\beta$ to give: $V_0 = 1 + \beta$. Typically, in CV-QKD protocols, we simply have $V = V_S + 1$, i.e., zero preparation noise ($\beta=0$). In this paper, we consider the effect of having non-zero preparation noise on Alice's mode preparation, i.e., $\beta>0$. We assume that this preparation noise cannot be controlled or manipulated by Eve.

In the analysis of CV-QKD protocols, the collective Gaussian
attacks~\cite{Nav06,Gar06,Pir08} are the most important.
In fact, up to a suitable symmetrization of the protocols~\cite{Ren09},
these attacks bound the most powerful eavesdropping strategy
allowed by quantum mechanics~\cite{Ren09}. The most general form of a
collective Gaussian attack is explicitly described in Ref.~\cite{Pir08}.
This consists in Eve interacting her (independent)
ancilla modes with Alice and Bob's mode for each run of the protocol
in such a way to generate a memoryless (one-mode) Gaussian channel.
Eve's ancillas are then collected in a quantum memory whose
measurement is optimized on Alice and Bob's classical
communications~\cite{Pir08}. For a practical implementation of the protocols,
the most important collective Gaussian attack is the one based on the
entangling cloner~\cite{Gross03} which is exactly the model considered in our paper.
This consists in Eve perfectly replacing the quantum channel between Alice and Bob with her own quantum channel where the loss is simulated by a beamsplitter with transmission $T$ (which ranges in value from $0$ to $1$). She then creates her ancilla modes which are  two-mode squeezed states~\cite{Ger05} (or commonly known as, Einstein-Podolsky-Rosen (EPR) states), with variance $W$. The modes of the EPR beam can be described by the operators $\hat{E}''$ and $\hat{E}$. She keeps one mode of the beam $\hat{E}''$ and injects the other mode $\hat{E}$ into the unused port of the beamsplitter, resulting in the output mode $\hat{E}'$. Eve then collectively detects all modes $\hat{E}'$ and $\hat{E}''$, gathered from each of the runs of the protocol, in a final coherent measurement. The final stages of the protocol consists in Alice and Bob publicly revealing a subset of their data in order to estimate the channel transmission $T$ and excess channel noise $W$~\cite{Sca09}. We also assume that Alice and Bob (and Eve) know the variance of the unknown preparation noise $\beta$ in order to properly estimate the channel noise as opposed to the sum of the channel noise and the preparation noise. However, the shot to shot displacement due to the excess preparation noise remains unknown to everyone. In the final steps of the protocol, Alice and Bob perform a reconciliation protocol~(e.g., see \cite{Cer07}) to correct any errors they might have between them and then finally privacy amplification~\cite{Sca09} to reduce Eve's knowledge of the key to a negligible, and safe amount.


%
\begin{figure}[!ht]
\begin{center}
\includegraphics[width=8cm]{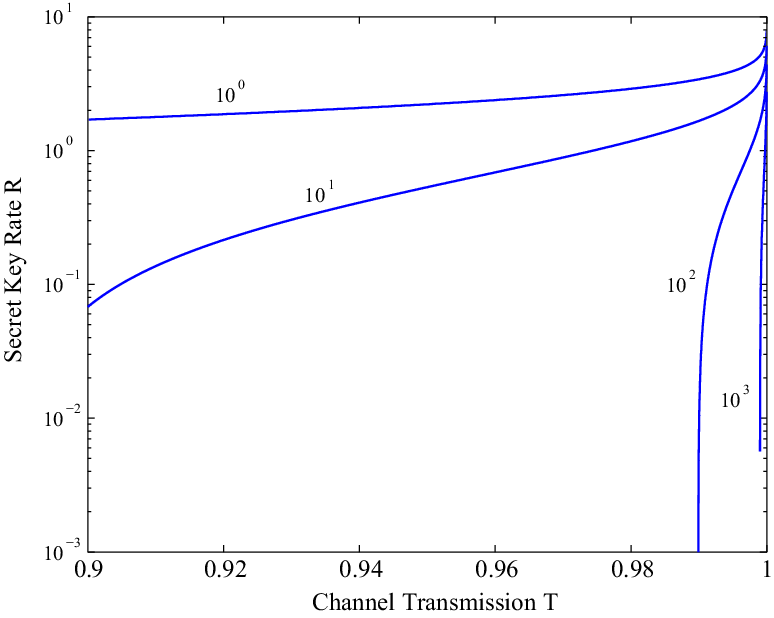}
\caption{Secret key rate $R^{\blacktriangleleft}$ versus channel transmission $T$ using reverse reconciliation. Increasing the amount of unknown classical noise on Alice's preparation modes in CV-QKD. Here the thermal radiation is increased: $V_0=1,10,10^2,10^3$ from left to right, where $W=1$~(lossy channel), $V_S=10^5$ and $V_0 =1$ is a pure vacuum mode.}\label{RR_Hom_distance1}
\end{center}
\end{figure}
\begin{figure}[!ht]
\begin{center}
\includegraphics[width=8cm]{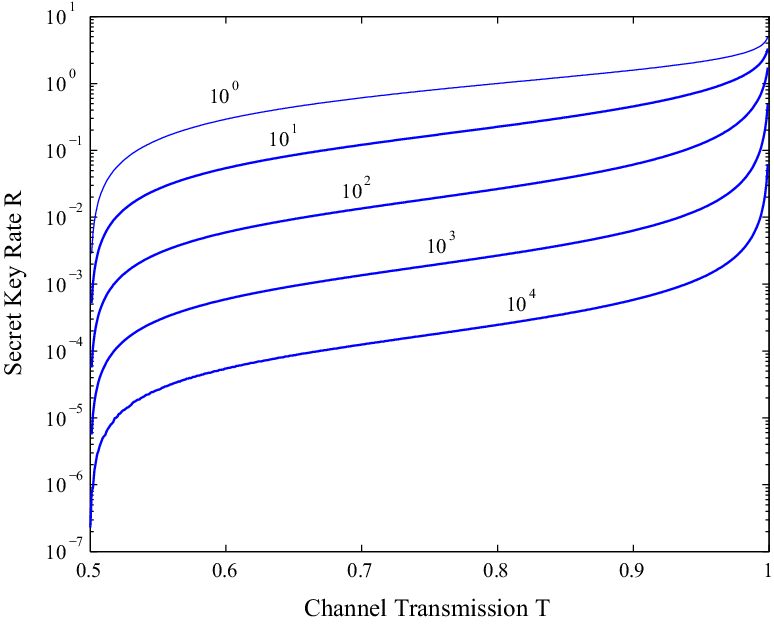}
\caption{Secret key rate $R^{\blacktriangleright}$ versus channel transmission $T$ using direct reconciliation. Increasing the amount of unknown classical noise on Alice's preparation modes in CV-QKD. Here the thermal radiation is increased: $V_0=1,10,10^2,10^3,10^4$ from top to bottom, where $W=1$, $V_S=10^5$. We find that direct reconciliation does not show any deterioration in channel loss when excessively large amounts of preparation noise is added.}\label{DR_Hom_distance2}
\end{center}
\end{figure}

\textbf{{\it Reverse Reconciliation}} - We begin our analysis by first using the CV-QKD protocol known as reverse reconciliation~\cite{F.Grosshans2003} which consists in Alice (and Eve) optimally estimating Bob's measurement outcomes. We note that the previous analysis given in~\cite{Fil08} also considered thermal state CV-QKD using
reverse reconciliation. However, for completeness, we give the derivation
for reverse reconciliation which will be helpful in calculating
the direct reconciliation case and for a comparison between the two protocols. The secret key rate $R^{\blacktriangleleft}$ for reverse reconciliation where Bob uses homodyne detection is given by $R^{\blacktriangleleft} := I(X_A:X_B) - I(X_B:E)$.
Here $I(X_A:X_B)$ is called the mutual information between Alice and Bob and defined in terms of the Shannon (or classical) entropy~\cite{Sha48}. The quantum mutual information between Eve and Bob $I(X_B:E)$ is given by the Holevo information~\cite{Hol73} and describes the most amount of information one can extract from a quantum state.

The secret key rate $R^{\blacktriangleleft}$ can be calculated~(see Appendix for complete derivation) for various values of preparation noise, i.e., $V_0 = 1,10,100,1000$. The results are plotted in Fig.~\ref{RR_Hom_distance1} for a lossy channel (i.e., $W=1$ which corresponds to Eve simply inserting a vacuum state into the unused port of the beamsplitter). We see that, as expected, the security is dependent on the channel transmission, and starts deteriorating rapidly as the excess preparation noise is increased. In fact, after only a modest increase in preparation noise (from $V_0=1$ to $V_0=10$), the secure region has shrunk to $T\approx > 0.89$.


\textbf{{\it Direct Reconciliation}} - We now turn our attention to another CV-QKD scheme known as direct reconciliation~\cite{F.Grosshans2002}. Direct reconciliation, was the first protocol to show that one could use Gaussian modulated coherent states to create a secure key. Unlike, reverse reconciliation, this protocol is a forward-way scheme where Bob (and Eve) are trying to optimally estimate the values of Alice's initial displacements, or encodings, $Q_A$ and $P_A$. However, direct reconciliation has the drawback in its inability to create a secret key when the loss is greater than $3$~dB. This corresponds to $T<0.5$ and can be intuitively thought of as Eve sharing more common information with Alice than Bob does. Consequently, reverse reconciliation (or post-selection~\cite{Sil02}) is usually considered the most practical CV-QKD protocol~\cite{Lev09}. However, as we will see, despite these shortcomings, direct reconciliation offers a surprising advantage as a potential platform for \textit{noise tolerant short-range QKD}.

The secret key rate $R^{\blacktriangleright}$ for direct reconciliation using homodyne detection is defined as $R^{\blacktriangleright} := I(X_A:X_B) - I(X_A:E)$
where $I(X_A:E)$ is again the Holevo quantity but now defined between Eve and Alice. We can now calculate the subsequent key rates~(see Appendix for details). In Fig.~\ref{DR_Hom_distance2} we have plotted the resulting secret key rates for various values of $V_0$ using $W=1$ and $V_S = 10^5$. We find that direct reconciliation has the amazing feature that as the preparation noise becomes more and more significant (even up to $10^4$ times that of the variance of the pure vacuum mode) only the secret key rate decreases and \textit{not} the channel transmission. So for any value of preparation noise the initial starting point is always $T=0.5$ (c.f. reverse reconciliation where modest increases in noise reduce the secure region close to unity transmission, i.e., see Fig.~\ref{RR_Hom_distance1}). The basic physics is that, for $T>0.5$, the presence of quantum noise always gives Alice and Bob a direct information advantage over Eve. Increased preparation noise reduces this advantage, but it always remains finite. In contrast, for reverse reconciliation, Alice's ability to estimate what Bob received is rapidly compromised by the preparation noise. This removes their information advantage over Eve.

\begin{figure}[!ht]
\begin{center}
\includegraphics[width=8cm]{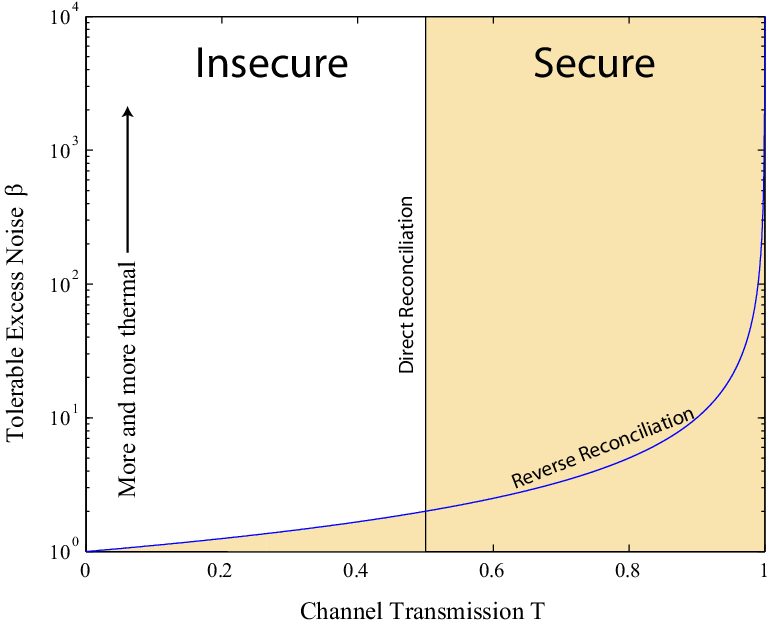}
\caption{Tolerable preparation (classical) excess noise $\beta = V_0-1$ versus channel transmission $T$ for direct and reverse reconciliation over a lossy channel. The area under the solid (blue) curve indicates the previous best secure region threshold using reverse reconciliation~\cite{Fil08}. However, for direct reconciliation, after $T=0.5$, one can immediately obtain many orders of magnitude improvement in the security threshold.}\label{excess_noise_best}
\end{center}
\end{figure}

In Fig.~\ref{excess_noise_best} we have a security threshold plot for direct and reverse reconciliation for $W=1$. The solid (blue) curve is the previous best bound derived using reverse reconciliation and is given by~\cite{Fil08}: $\beta < (1-T)^{-1}$. On the same plot we have the new direct reconciliation bound which shows a substantial improvement over the previous reverse reconciliation bound. Remarkably, we can see how direct reconciliation is unaffected by the channel transmission once $T>0.5$ and is secure for a \textit{minimum} of 4 orders of magnitude of preparation noise. Therefore, it is best to use reverse reconciliation when $T\leq 0.5$ and direct reconciliation when $T>0.5$. Additionally, this result is robust to the addition of small amounts of excess noise on the quantum channel (i.e., $W>1$) which moves the transmission limit slightly over $50\%$ but retains qualitatively the same behavior as the lossy case~\cite{Wee10}.


%
%


%
\begin{figure}[!ht]
\begin{center}
\includegraphics[width=8cm]{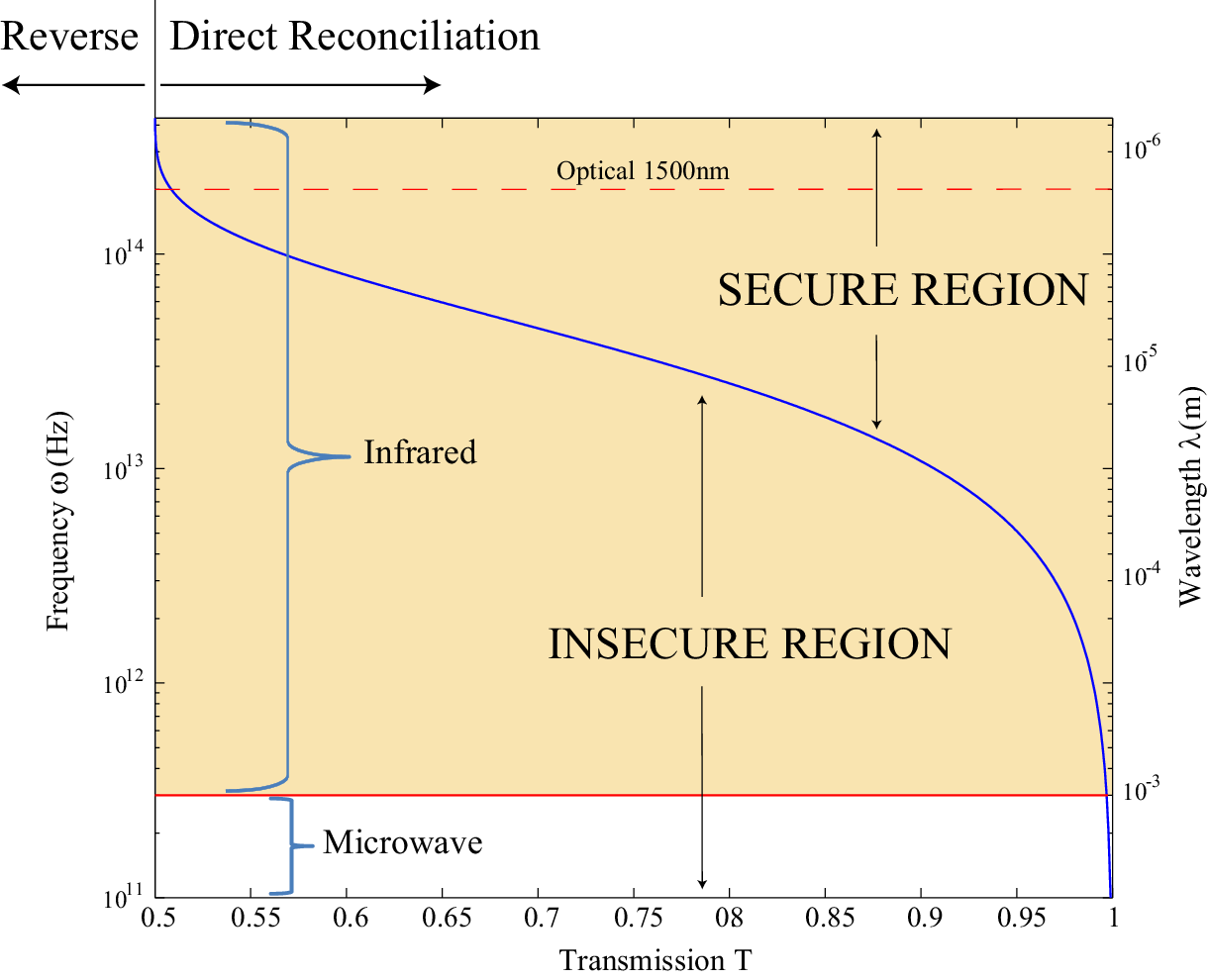}
\caption{Security of quantum cryptography over various electromagnetic wave frequencies (at room temperature) as a function of channel transmission. Moving our way from the infrared spectrum ($430$~THz) and into the microwave spectrum ($300$~GHz). Our results show that direct reconciliation should be used when channel losses are less than $50\%$ and reverse reconciliation otherwise. We note that at each point the same impurity applies to both Alice and Eve with $V_S=10^8$.}\label{DR_microwave3}
\end{center}
\end{figure}

\textbf{{\it Infrared to Microwave Quantum Cryptography}} - It is interesting to consider a possible application of our results: wireless CV-QKD at infrared to microwave frequencies. Today, a large number of popular wireless communication technologies rely on such frequencies to distribute information. Due to the ubiquitous nature of such devices, their security is of fundamental importance. Moving to frequencies lower than optical rules out discrete variable QKD because of the lack of photon counting capabilities. The problem for CV-QKD is that operating at lower frequencies at room temperature inevitably introduces a significant amount of thermal noise. In contrast to the previous section, we now consider a simplified wireless communication protocol where \textit{both} Alice's preparation modes and the quantum channel (Eve) are affected by the thermal background. When considering Eve we assume that she prepares her attack within a cryostat which allows her to essentially prepare pure modes away from the effect of the background radiation. Then to cover her tracks she adds \textit{known} excess noise to her pure states to emulate the thermal noise of the environment.

In the previous section we showed that direct reconciliation is significantly more robust against preparation noise than reverse reconciliation and is consequently better suited to our current analysis. Given that, the next step is to calculate how strong the thermal modes are at particular frequencies from optical down to the microwave ($1$~GHz ($\lambda=30$~cm) to $300$~GHz ($\lambda=1$~mm)). To do this we first write the average photon number $\bar{n}$ in terms of the quadrature variance $V$ using $\bar{n} = \langle \hat{a}^{\dagger} \hat{a} \rangle =  (V-1)/2 \Longrightarrow V = 2 \bar{n}+1$
where we have symmetrized both quadratures, i.e., $V := V(\hat{Q}) = V(\hat{P})$ and the annihilation operator $\hat{a}$ is defined as $\hat{a} = (\hat{Q} + i \hat{P})/2$. Secondly, the average photon number is equal to $\bar{n} = [\exp(\hbar \omega / k_B T) - 1]^{-1}$~\cite{Ger05}
and represents the blackbody radiation spectrum. For example, at room temperature $T=300$~K and using a microwave frequency of $\omega=1$~GHz we find that the variance of the thermal mode is $V=7.85 \times 10^4$; while at the other end of the microwave spectrum ($\omega=300$~GHz) the variance is $V=2.63 \times 10^2$.

Using the analysis from the previous section, we can calculate the secret key rates using direct reconciliation. In Fig.~\ref{DR_microwave3} we plot the security of CV-QKD from the optical frequency ($1550$~nm) into the infrared region and down into the microwave frequency as a function of channel line transmission. We point out that the secure region corresponds to $R > 0$. We find a window of security for CV-QKD throughout all of the infrared region and into the microwave frequency albeit with smaller allowed levels of loss. In the mid-infrared region transmission of $T \approx 0.8$ is required whilst in the case of the microwave region we see that a secure key can only be generated when the transmission is higher than $T \approx 0.9969$. Nonetheless, it is interesting that a small security window, in principle, exists. Future analysis will look at improving the region where infrared and microwave CV-QKD is secure. For example, in \cite{S.Pirandola2008} they showed that the security thresholds for direct reconciliation could be improved (and in fact beat the 3~dB loss limit) if two-way quantum communication was used. Furthermore, post-selection~\cite{Sil02} could also be used to investigate a possible way to combat the high preparation noise.

\textbf{{\it Conclusion}} - In conclusion, we have shown that when considering unknown preparation noise in continuous-variable QKD, direct reconciliation is significantly more robust than reverse reconciliation when the channel loss does not exceed $50\%$. Incredibly, direct reconciliation showed no deterioration in the loss threshold, only in secret key rates, even when the variance of the thermal noise is as much as $10^4$ times greater than that of the pure vacuum mode. Furthermore, we have shown that infrared to microwave quantum cryptography is, in principle, possible over short distances when using continuous variables and opens up the possibility of further avenues of investigations. In conclusion, we have shown that when considering unknown preparation noise in continuous-variable QKD, direct reconciliation is significantly more robust than reverse reconciliation when the channel loss does not exceed $50\%$. Incredibly, direct reconciliation showed no deterioration in the loss threshold, only in secret key rates, even when the variance of the thermal noise is as much as $10^4$ times greater than that of the pure vacuum mode. Furthermore, we have shown that infrared to microwave quantum cryptography is, in principle, possible over short distances when using continuous variables and opens up the possibility of further avenues of investigations.

Acknowledgments -- C.~W. and T.~R. would like to thank the Australian Research Council (ARC) for funding. S.~P. was supported by a Marie Curie Action of the European
Community. S.~L. and C.~W. would like to thank the W. M. Keck Foundation Center for Extreme Quantum Information Theory for funding. C.W. would also like to thank Vladyslav Usenko, Radim Filip, Warwick Bowen, Nathan Walk, Denis Sych, Franco Wong, Travis Humble and Jeff Shapiro for helpful discussions. We would also like to thank Nathan Walk, Carlo Ottaviani and Xiang-Chun Ma for pointing out a mistake in a previous version.

\section{Appendix}

\subsection{Introduction to Gaussian Formalisms}

Here we introduce some of the Gaussian tools and techniques required for our analysis. Such Gaussian formalisms can be found elsewhere in the literature typically in the context of \textit{quantum information using continuous variables} (for example, see \cite{Fer05,Ade06,Gar07,S.Pirandola2008}). However, in what follows, we give the reader a self-contained treatment for what is needed to understand and derive the results given in the main text of the paper.

To begin with, we can define the quadrature row vector $\hat{\textbf{Y}}$, which describes a bosonic system~\cite{Bra05} of $n$ modes, as:
\begin{align}
\hat{\textbf{Y}} = (\hat{Q}_1, \hat{P}_1, ..., \hat{Q}_n, \hat{P}_n),
\end{align}
where $\hat{Y}_l$ is the $l$th element of the vector. This satisfies the commutator relation:
\begin{align}\label{Eq: commutator}
[\hat{Y}_l, \hat{Y}_m] = 2i \Omega_{lm},
\end{align}
for $1 \leq l,m, \leq 2n$. Here the matrix ${\bf \Omega}$ defines the symplectic form and is given as
\begin{align}
{\bf \Omega}:= \bigoplus_{k=1}^{n} \left(
                                     \begin{array}{cc}
                                       0 & 1 \\
                                       -1 & 0 \\
                                     \end{array}
                                   \right),
\end{align}
where $\Omega_{lm}$ denote the row $l$ and column $m$ of the matrix, e.g., $\Omega_{11}$ is the first row and first column entry of $\Omega$. Firstly, the notation for the above matrix can be explained with a simple example. For $n=2$ mode case the direct sum $\bigoplus$ means that we form two $2 \times 2$ block diagonal matrices to create a larger $4 \times 4$ matrix, i.e.,
\begin{align}
{\bf \Omega}_{n=2}:= \bigoplus_{k=1}^{2} \left(
                                     \begin{array}{cc}
                                       0 & 1 \\
                                       -1 & 0 \\
                                     \end{array}
                                   \right)
= \left(
  \begin{array}{cc|cc}
    0 & 1 &  &  \\
    -1 & 0 &  &  \\\hline
     &  & 0 & 1 \\
     &  & -1 & 0 \\
  \end{array}
\right),
\end{align}
where the empty spaces indicate zero elements. Secondly, the compact version of the commutator given in Eq.~(\ref{Eq: commutator}) can now be understood using the simple case of one mode ($n=1$) where
\begin{align}
{\bf \Omega}_{n=1}:= \left(
                                     \begin{array}{cc}
                                       0 & 1 \\
                                       -1 & 0 \\
                                     \end{array}
                                   \right),
\end{align}
and $\hat{Y}_1 := \hat{Q}_1$ and $\hat{Y}_2 := \hat{P}_1$ for $l,m = 1,2$. Therefore Eq.~(\ref{Eq: commutator}) is a compact way of saying the following:
\begin{align}\nonumber
[\hat{Y}_1,\hat{Y}_1] &= [\hat{Q}_1,\hat{Q}_1] = 2i \Omega_{11} = 0,\\\nonumber
[\hat{Y}_1,\hat{Y}_2] &= [\hat{Q}_1,\hat{P}_2] = 2i \Omega_{12} = 2i,\\
[\hat{Y}_2,\hat{Y}_1] &= [\hat{P}_2,\hat{Q}_1] = 2i \Omega_{21} = -2i,\\\nonumber
[\hat{Y}_2,\hat{Y}_2] &= [\hat{P}_2,\hat{P}_2] = 2i \Omega_{22} = 0.
\end{align}

A Gaussian bosonic state $\rho$ is fully characterized by its displacement
\begin{align}
\langle \hat{\textbf{Y}} \rangle = {\rm Tr} (\hat{\textbf{Y}}  \rho),
\end{align}
and its correlation matrix (CM). The various elements of a correlation matrix $\textbf{V}$ can be calculated using the following formulas. Firstly, the off-diagonal terms:
\begin{align}
V_{lm}:= \frac{1}{2} \langle \hat{Y}_l \hat{Y}_m + \hat{Y}_m \hat{Y}_l \rangle - \langle \hat{Y}_l \rangle  \langle \hat{Y}_m \rangle,
\end{align}
and the diagonal elements:
\begin{align}
V_{ll}= \langle \hat{Y}_l^2 \rangle - \langle \hat{Y}_l \rangle^2 := V(\hat{Y}_l).
\end{align}
Qualitatively it means that the diagonal terms contain the variances whilst the off-diagonal terms contain the correlations. An example might help here. A simple vacuum state, where the variance of the quadratures is normalized to one, has a CM given by
\begin{align}
\textbf{V}_{vac} = \left(
  \begin{array}{cc}
    1 & 0 \\
    0 & 1 \\
  \end{array}
\right).
\end{align}
We can see that the variance of each quadrature is on the diagonal, whilst on the off-diagonal we have the zero correlation (meaning they are independent) terms.

The von Neumann entropy~\cite{Nie00}
\begin{align}
S(\rho) = -{\rm Tr}(\rho {\rm log}_2 \rho),
\end{align}
of a Gaussian state $\rho$ can be written in terms of its \textit{symplectic eigenvalues} $\nu_k$ as \cite{Hol99}
\begin{align}\label{eq: von neuman entropy}
S(\rho) = \sum_{k=1}^{n} g (\nu_k),
\end{align}
where
\begin{align}
g(\nu) := \Big(\frac{\nu+1}{2}\Big) {\rm log}_2 \Big(\frac{\nu+1}{2}\Big) - \Big(\frac{\nu-1}{2}\Big) {\rm log}_2 \Big(\frac{\nu-1}{2}\Big).
\end{align}
These symplectic eigenvalues can be calculated using the formula
\begin{align}\label{Eq: nu}
\nu = |i {\bf \Omega V}|,
\end{align}
where $\nu \geq 1$. The above notation means that you first find the eigenvalues of the matrix $i {\bf \Omega V}$ and then take the absolute values. As it turns out these eigenvalues (known also as the symplectic spectrum) are a powerful tool which allows one to determine many important features of a Gaussian system. Although Eq.~(\ref{Eq: nu}) gives one a way of calculating the spectrum, the output from using such a formula can sometimes lead very complicated equations. In certain circumstances, we are able to simplify the calculation of the eigenvalues. Let's look at that now. First, consider a generic two-mode CM in block form
\begin{align}\label{eq: CM 2 mode}
\mathbf{V}= \left(
              \begin{array}{cc}
                \mathbf{A} & \mathbf{C} \\
                \mathbf{C}^T & \mathbf{B} \\
              \end{array}
            \right).
\end{align}
It is known \cite{Alex} that its symplectic eigenvalues $\nu _{1}$ and $\nu _{2}$ can be written in the form
\begin{equation}\label{eq: symplectic eigenvalues def}
\nu _{1,2}=\sqrt{\frac{1}{2}\Big(\Delta \pm \sqrt{\Delta ^{2}-4\det \mathbf{V}}\Big)}~,
\end{equation}
where $\det \mathbf{V}$ means the determinant of the matrix $\textbf{V}$ and
\begin{align}
\Delta :=\det \mathbf{A}+\det \mathbf{B}+2\det \mathbf{C}.
\end{align}
In particular, let us consider a CM of the form
\begin{align}
\mathbf{V}= \left(
              \begin{array}{cc}
                a\mathbf{I} & \sqrt{T}c\mathbf{Z} \\
                \sqrt{T}c\mathbf{Z} & b\mathbf{I} \\
              \end{array}
            \right),
\end{align}
where $c\geq 0$, $T\in \lbrack 0,1]$ and
\begin{align}
\mathbf{I} = \left(
               \begin{array}{cc}
                 1 & 0 \\
                 0 & 1 \\
               \end{array}
             \right)
\hspace{.5cm}
\mathbf{Z} = \left(
               \begin{array}{cc}
                 1 & 0 \\
                 0 & -1 \\
               \end{array}
             \right).
\end{align}
We can easily verify that $\det
\mathbf{V}=(ab-c^{2}T)^{2}$ and $\Delta =a^{2}+b^{2}-2c^{2}T$. As a consequence, the eigenvalues take the simple expression
\begin{equation}\label{Spectrum_simple1}
\nu _{1,2}:= \frac{1}{2} \Big(\sqrt{y} \pm (a-b)\Big) ~,
\end{equation}
where $y:=(a+b)^{2}-4c^{2}T\geq 4$.

\subsection{Reverse Reconciliation}

The secret key rate $R^{\blacktriangleleft}$ for reverse reconciliation where Bob uses homodyne detection is given by
\begin{align}\label{eq: secret key rate hom RR}
R^{\blacktriangleleft} := I(X_A:X_B) - I(X_B:E),
\end{align}
where $I(X_A:X_B)$ is known as the mutual information between Alice and Bob and $I(X_B:E)$ the mutual information between Bob and Eve. We have a secure key when the key rate is positive, i.e.,
\begin{align}
R > 0.
\end{align}
Another way to think about this is a secure key can be synthesized when Eve has less information than Alice and Bob:
\begin{align}
I(X_A:X_B) > I(X_B:E).
\end{align}
We also note that, for Alice and Bob, the variable $\hat{X}$ corresponds to either of the two quadratures $\{\hat{Q},\hat{P}\}$, such that
\begin{align}
\hat{X}_A &= \{\hat{Q}_A,\hat{P}_A\},\\
\hat{X}_B &= \{\hat{Q}_B,\hat{P}_B\}.
\end{align}
Firstly, let us calculate the mutual information between Alice and Bob where
\begin{align}\label{eq: alice and bob mutual Hom RR}
I(X_A:X_B):= H(X_B) - H(X_B|X_A).
\end{align}
Here
\begin{align}\label{eq: HXB}
H(X_B) = \frac{1}{2} \log_2 V(\hat{X}_B),
\end{align}
is the Shannon (or classical) entropy and
\begin{align}\label{eq: HXBXA}
H(X_B|X_A) = \frac{1}{2} \log_2 V(\hat{X}_B|X_A),
\end{align}
is known as the conditional Shannon entropy \cite{Sha48}. To determine $\hat{X}_B$ we set up a generic quantum channel with transmission $T \in [0,1]$ with excess noise $\hat{N}$ and model it using a beamsplitter equation, where the transmitted output (received by Bob) is given by:
\begin{align}
\hat{X}_B = \sqrt{T} \hat{X}_A + \sqrt{1-T} \hat{N}.
\end{align}
The variance of the above equation is given by
\begin{align}\label{eq: Bob variance}
V(\hat{Q}_B) &= V(\hat{P}_B) = (1-T) W + T V := b_V,
\end{align}
where both quadratures have been symmetrized and $V(\hat{N}):= W$. Also, the variance of Alice's modes is given by
\begin{align}
V = V_S + V_0,
\end{align}
where $V_S$ is the variance of the initial signal encodings and $V_0$ is the variance of the vacuum state (see main text for more detail). In Eq.~(\ref{eq: HXBXA}) the conditional variance term $V(\hat{X}_B|X_A)$ is derived by setting up an optimal estimator equation (e.g., see its use in \cite{S.Pirandola2008} or \cite{Wee06}):
\begin{align}
V(\hat{X}_B|X_A) = V(\hat{X}_B) - \frac{|\langle \hat{X}_B X_S \rangle|^2}{V(X_S)},
\end{align}
where specifically here we have $X_S \in \{Q_A,P_A\}$ (the signal) rather than $\hat{X}_A$ (signal plus noise) because it is Bob's estimate of Alice's signal not his estimate of both the signal and noise. Calculating this explicitly we get:
\begin{align}\label{eq: Bob cond var}
V(\hat{Q}_B|Q_A) = V(\hat{P}_B|P_A) = (1-T) W + T V_0 := b_1.
\end{align}
Using Eq.~(\ref{eq: alice and bob mutual Hom RR}) with Eqs.~(\ref{eq: HXB}) and (\ref{eq: HXBXA}) we calculate Alice and Bob's mutual information to be
\begin{align}\label{eq: alice and bob mutual hom RR formula}
I(X_A:X_B) = \frac{1}{2} \log_2 \Big[\frac{(1-T) W + T V_S + T V_0}{(1-T) W + T V_0}\Big].
\end{align}

We now turn our attention to calculating the mutual information between Eve and Bob. This is given by the Holevo information \cite{Hol73} defined as
\begin{align}
I(X_B:E) := S(E) - S(E|X_B).
\end{align}
In the literature it is also common to use the notation $\chi$ for the Holevo (information) bound (for more background on both the classical and quantum information formulas, see e.g., \cite{Cov06,Nie00}).

Here the quantum entropies $S(E):= S(\rho_E)$ and $S(E|X_B)$ are found by calculating the eigenvalues, or symplectic spectrum $\nu$, of their corresponding CMs: $\textbf{V}_E$ and $\textbf{V}_{E|X_B}$, respectively.
%
Eve's CM is made up from the two modes $\hat{E}'$ and $\hat{E}''$ (see main text for details) and is given by
\begin{align}\label{eq: eves CM}
\textbf{V}_{E} (V,V) = \left(
              \begin{array}{cc}
                {\bf \Delta}[e_{V},e_V] & \varphi \textbf{Z} \\
                \varphi \textbf{Z} & W \textbf{I} \\
              \end{array}
            \right),
\end{align}
where
\begin{align}
e_V:= (1-T) V + T W,
\end{align}
and the notation ${\bf \Delta}[\cdot,\cdot]$ simply means a diagonal matrix with the arguments $[\cdot,\cdot]$ on the diagonal entries and also
\begin{align}
\varphi = [T(W^2 -1)]^{1/2}.
\end{align}
%
Eve's symplectic spectra can be determined by using Eq.~(\ref{Spectrum_simple1})
\begin{align}
\nu_E = \frac{1}{2} [\sqrt{(e_V+W)^2 - 4 T (W^2 -1)} \pm (e_V - W)].
\end{align}
Eve's conditional CM is given by
\begin{align}
\textbf{V}_{E|X_B} =\textbf{V}_{E} - (\beta_V)^{-1} \textbf{C} {\bf \Pi} \textbf{C}^T,
\end{align}
where $\textbf{V}_E$ is defined in Eq.~(\ref{eq: eves CM}) and
\begin{align}
{\bf \Pi} := \left(
               \begin{array}{cc}
                 1 & 0 \\
                 0 & 0 \\
               \end{array}
             \right).
\end{align}
Furthermore, $\textbf{C}$ is a $4 \times 2$ matrix describing the quantum correlations between Eve's modes $\{\hat{E}',\hat{E}''\}$ and Bob's output mode $\hat{X}_B$ and is defined as
\begin{align}
\textbf{C} := \left(
               \begin{array}{c}
                 \langle \hat{E}' \hat{X}_B \rangle \textbf{I} \\
                 \langle \hat{E}'' \hat{X}_B \rangle \textbf{Z} \\
               \end{array}
             \right)
             =
             \left(
               \begin{array}{c}
                 \xi \textbf{I} \\
                 \phi \textbf{Z} \\
               \end{array}
             \right),
\end{align}
where
\begin{align}
\xi = -\sqrt{T (1-T)} (V_S + V_0 - W),
\end{align}
and
\begin{align}
\phi = \sqrt{1-T} \sqrt{W^2-1},
\end{align}
and we have used
\begin{align}
\hat{X}_B = \sqrt{T} \hat{X}_A + \sqrt{1-T} \hat{E},
\end{align}
and
\begin{align}
\hat{E}' = -\sqrt{1-T} \hat{X}_A +\sqrt{T} \hat{E}.
\end{align}
Using the above we find that Eve's conditional CM $\textbf{V}_{E|X_B}$ has the form
%
\begin{align}
\textbf{V}_{E|X_B} =
\left(
\begin{array}{cc}
\mathbf{A} & \mathbf{C} \\
\mathbf{C}^{T} & \mathbf{B}%
\end{array}%
\right),
\end{align}
where
\begin{align}
\mathbf{A} &=
\left(
  \begin{array}{cc}
    \frac{VW}{T (V-W) + W} & 0 \\
    0 & (1-T) V + TW \\
  \end{array}
\right),\\\nonumber
\mathbf{B} &=
\left(
  \begin{array}{cc}
    \frac{1-T + TWV}{TV + W - TW} & 0 \\
    0 & W \\
  \end{array}
\right),\\\nonumber
\mathbf{C} &=
\left(
  \begin{array}{cc}
    \sqrt{T (W^2-1)} \Big[\frac{V}{TV + W - T W}\Big] & 0 \\
    0 & -\sqrt{T (W^2-1)} \\
  \end{array}
\right).
\end{align}
%
Using Eq.~(\ref{eq: symplectic eigenvalues def}) the corresponding symplectic spectra $\nu_{E|X_B}$ of $\textbf{V}_{E|X_B}$ can be calculated. 
The final secret key rate $R^{\blacktriangleleft}$ can now be calculated numerically, using Eq.~(\ref{eq: secret key rate hom RR}) with the appropriate formulas, for various values of preparation noise.

\subsection{Direct Reconciliation}

The secret key rate $R^{\blacktriangleright}$ for direct reconciliation using homodyne detection is given by
\begin{align}\label{eq: secret key rate hom DR}
R^{\blacktriangleright} := I(X_A:X_B) - I(X_A:E),
\end{align}
where $I(X_A:X_B)$ has already been calculated in Eq.~(\ref{eq: alice and bob mutual hom RR formula}). For Eve, we have
\begin{align}
I(X_A:E) := S(E) - S(E|X_A),
\end{align}
where again we have already calculated $S(E)$ previously and $S(E|X_A)$ is calculated from the spectrum of the conditional CM $\textbf{V}_{E|X_A}$. Eve's conditional CM for homodyne detection using direct reconciliation is equal to 
\begin{align}
\textbf{V}_{E|Q_A} = \textbf{V}_E (V_0,V),
\end{align}
where $\textbf{V}_E$ is defined in Eq.~(\ref{eq: eves CM}). Using Eq.~(\ref{Eq: nu}) the corresponding symplectic spectra $\nu_{E|X_A}$ is:
\begin{align}
\nu_{E|X_A} = \frac{1}{\sqrt2} \Big(\sqrt{|F \pm \sqrt{G}|}\Big)
\end{align}
where
\begin{align}\nonumber
F &= V V_0 +T(2+(T-2)V V_0) - TW(T-1)(V+V_0)\\ &+ W^2(T-1)^2,
\end{align}
and
\begin{align}\nonumber
G &= (T-1)^2 (T^2(V-W)^2(V_0-W)^2+(-V_0V+W^2)^2\\&+2T(V-W)(W-V_0)(-2+V V_0 +W^2)).
\end{align}
The final secret key rate $R^{\blacktriangleright}$ can now be calculated numerically, using Eq.~(\ref{eq: secret key rate hom DR}) with the appropriate formulas, for various values of preparation noise.

\end{document}